\documentstyle[aps,epsf,pre,preprint]{revtex}

\draft
\tightenlines
\begin{document}

\title{Effect of Subphase Ca$^{++}$ Ions on the Viscoelastic
Properties of Langmuir Monolayers}

\author{R. S. Ghaskadvi, Sharon Carr, and Michael Dennin}
\address{Department of Physics and Astronomy}
\address{University of California at Irvine}
\address{Irvine, CA 92697-4575.}

\date{\today}

\maketitle

\begin{abstract}

It is known that the presence of cations like Ca$^{++}$ or Pb$^{++}$
in the water subphase alters the pressure-area isotherms for fatty
acid monolayers. The corresponding lattice constant changes have been
studied using x-ray diffraction. Reflection-absorption spectroscopy
has been used to probe the chemical composition of the film. 
We report on the first
measurements of the time evolution of the shear viscosity of arachidic
acid monolayers in the presence of Ca$^{++}$ ions in the subphase. We find
that the introduction of Ca$^{++}$ ions to the water subphase results in an
increase of the film's viscosity by at least three orders of magnitude. 
This increase occurs
in three distinct stages. First, there is a rapid change in the viscosity of
up to one order of magnitude. This is followed
by two periods, with very different time constants, of a relatively
slow increase in the viscosity over the next 10 or more hours. The 
corresponding time constants for this rise decrease as either the
subphase pH or Ca$^{++}$ concentration is increased.
 
\end{abstract}


\section{Introduction}

Over the last ten years there has been a renewed interest in the
study of Langmuir monolayers \cite{Mono},  due to the development
and application of a number of
powerful tools like x-ray diffraction \cite{DPLKPGE87,KHLM87},
Brewster Angle microscopy (BAM) \cite{HM91,HM91b}, and fluorescence
microscopy \cite{SK93,RHMSTK94,MKAR90}.
Langmuir monolayers are monomolecular films at
the air-water interface formed by amphiphilic molecules.
Typically, these molecules have a long hydrophobic chain oriented away 
from the water
surface and a polar, hydrophilic headgroup that interacts with
the components of the aqueous subphase. Langmuir monolayers
serve as an excellent model for biological membranes and for
surfactant stabilizers that are added to foams. Also, they are the
starting point for Langmuir-Blodgett depositions where a solid
substrate is passed through the Langmuir monolayers, transferring
one or more layers of the molecules. For all three of these applications,
understanding the interaction between Langmuir monolayers and ions in the
subphase is important for two reasons. First, the ions are often naturally
present in these systems, either as biologically relevant chemicals or as
contaminants. Second, the ions provide a mechanism for controlling the
mechanical properties of the films, which is especially important in
applications related to foams and Langmuir-Blodgett depositions. 

A number of techniques, including pressure-area
isotherms \cite{MAXK82,PCBV90,BBGM91},
reflection-absorption spectrometry \cite{GH94},
and x-ray diffraction \cite{SBMZD92},
have been used to study the effects of divalent cations on the monolayer
structure. These studies have highlighted the important role played
by pH, especially for fatty acid monolayers, in modifying the
effects of divalent cations on the structure of the monolayer.
The equilibrium phase behavior of fatty acids on a pure water, or
low pH subphase, have been extensively studied \cite{Phase}.
There is a generally
applicable phase diagram that consists of both ``tilted'' and 
``untilted'' phases. A tilted phase is one in which the monolayer
tails are tilted with respect to the surface normal. Generally, the
untilted phases occur at higher pressures. One of the 
main effects of the calcium ions, as the pH is increased, is to
lower the transition pressure between the various phases \cite{SBMZD92}.
Ultimately, at very high pH, the tilted phases no longer appear to exist.
This lowering of the transition pressure is often referred to as
a ``stiffening'' of the monolayer. A common feature
of these studies is that no long term variations
in the monolayer properties were measured. This is reasonable if
chemical equilibrium with the ions in 
solution is reached relatively rapidly.

Despite the evidence from pressure-area isotherms that the cations cause a
stiffening of the monolayer, there has been minimal efforts to
measure effects of cations on the viscoelastic properties of the
monolayer \cite{BGR88,YYZ90}.
In this paper, we report on a series of measurements of the viscoelastic
properties of arachidic acid monolayers in the presence of Ca$^{++}$.
We have looked at the effect of pH and Ca$^{++}$ concentration on the time
evolution of three properties of the monolayer:
the isotherms; the viscosity ($\eta$); and the complex shear 
modulus G. 
Our isotherm results at t = 1 hr are consistent with previous measurements
of fatty acids and divalent cations \cite{SBMZD92}. However, we have found
a slow change in the viscoelastic properties of the monolayer
over a long time period. This behavior suggests interesting kinetics for
the chemical reaction between the arachidic acid
and the Ca$^{++}$.

\section{Experimental Details}
\label{exp}

\begin{figure}
\epsfxsize = 3.5in
\centerline{\epsffile{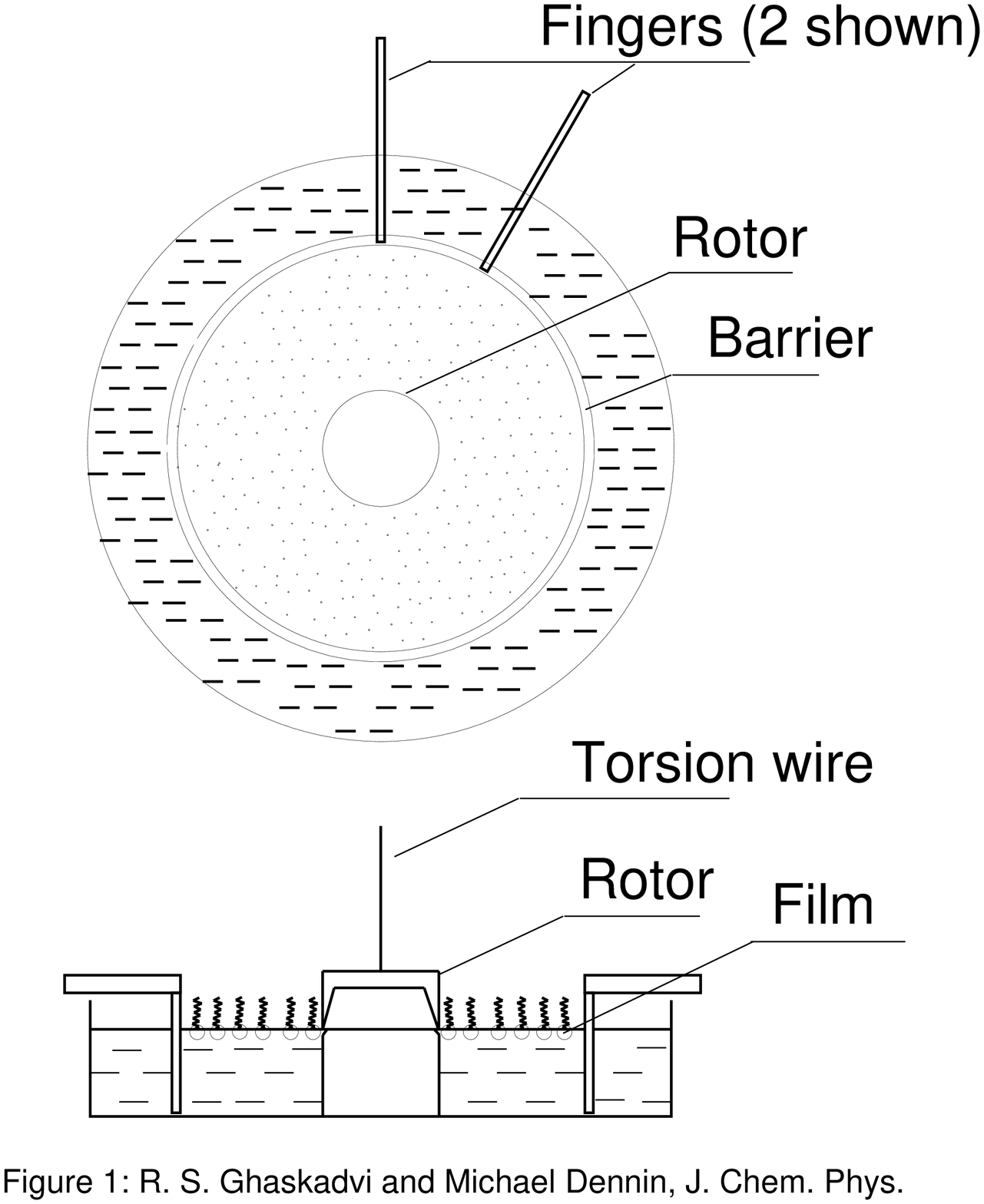}}
\caption{\label{app} Schematic drawing of the apparatus.}
\label{fig:app}
\end{figure}

The viscoelastic properties were measured using a two-dimensional
Couette viscometer that is described in detail elsewhere \cite{GD98}. A
schematic of the apparatus is given in Fig.~\ref{app}.
A circular barrier made of twelve individual teflon fingers
is immersed into water in a circular trough. A circular knife-edge
torsion pendulum (rotor) hangs by a wire so that it just touches the water
surface in the center of the trough. A stationary teflon disk is placed
in the water just under the pendulum. The disk has the same diameter as 
the knife-edge pendulum.
A Langmuir monolayer is made at the annular air-water interface
between the barrier and the rotor knife-edge. The barrier can be 
compressed or expanded to control the monolayer pressure and
rotated to generate a two dimensional Taylor-Couette flow.
The angular position of the rotor can be
measured by means of a pick-up coil attached to the rotor. This is used
to measure the torque generated by flow in the monolayer on the inner
rotor. The torque provides a measurement of the monolayer
viscosity.
In addition, an external torque can be applied to the rotor
by manipulating an external magnetic field. This allows for both
oscillatory measurements of the linear shear response of the monolayer
and measurements of stress relaxation curves for monolayers.

The apparatus is also equipped with a Brewster Angle Microscope (BAM) for
observation of the domain structure of the film. The BAM image
measures the relative reflectivity of p-polarized light
incident on the monolayer at the Brewster angle for pure water.
Variations of reflectivity of the monolayer correspond to changes
in the orientation of the tilted molecules from domain to domain. 

To study the effect of cations, it is imperative to start with water
that has minimal ionic content. We achieved this by passing de-ionized
water through a Millipore filter to obtain water with resistivity in
excess of $18 \  {\rm M}\Omega{\rm /cm}$. The concentration of Ca$^{++}$
was set by adding CaCl$_{2}$.2H$_{2}$O to the purified water. Most of
the experiments used a 0.65 mM Ca$^{++}$ concentration so that the
results would be comparable with Ref.~\cite{SBMZD92}.

The arachidic acid monolayer was made from a chloroform solution. The
solution was placed on the aqueous subphase with a microsyringe and
allowed to relax for about 20 minutes to facilitate the evaporation of the
solvent. Then it was compressed to the pressure of 9 dyne/cm.
All the data presented here were taken at 22 $^\circ$C. At this temperature
and pressure, the monolayer is in the L$_2$ phase.  One hour after
the solution was placed on the subphase, the equilibrium angle,
$\theta_1$, of the rotor was measured.
The outer barrier was set into
rotation to generate a Couette flow. The Couette flow causes a torque $\tau$ 
on the rotor displacing it to a new equilibrium position $\theta_2$
such that  $\tau = \kappa (\theta_2 - \theta_1)$, where
$\kappa$ is the torsion constant of the wire. After rotating the barrier for
about 5 minutes to achieve equilibrium, $\theta_2$ was measured, and then
the rotation was stopped. 
This series of experiments was repeated every hour.
The isotherms as well as the complex shear modulus were measured separately.
The barrier rotation 
rate was
0.0237 rad/sec. With $R_{\rm inner} = 3.81 \ {\rm cm}$ and
$R_{\rm barrier} = 6.5 \ {\rm cm}$, this corresponds 
to a shear rate of $0.057 \ s^{-1}$. G was measured at $\omega$ = 
0.251 rad/sec.

\section{Results} 
\label{res}

\begin{figure}
\epsfxsize = 3.5in
\epsfysize = 3.5in
\centerline{\epsffile{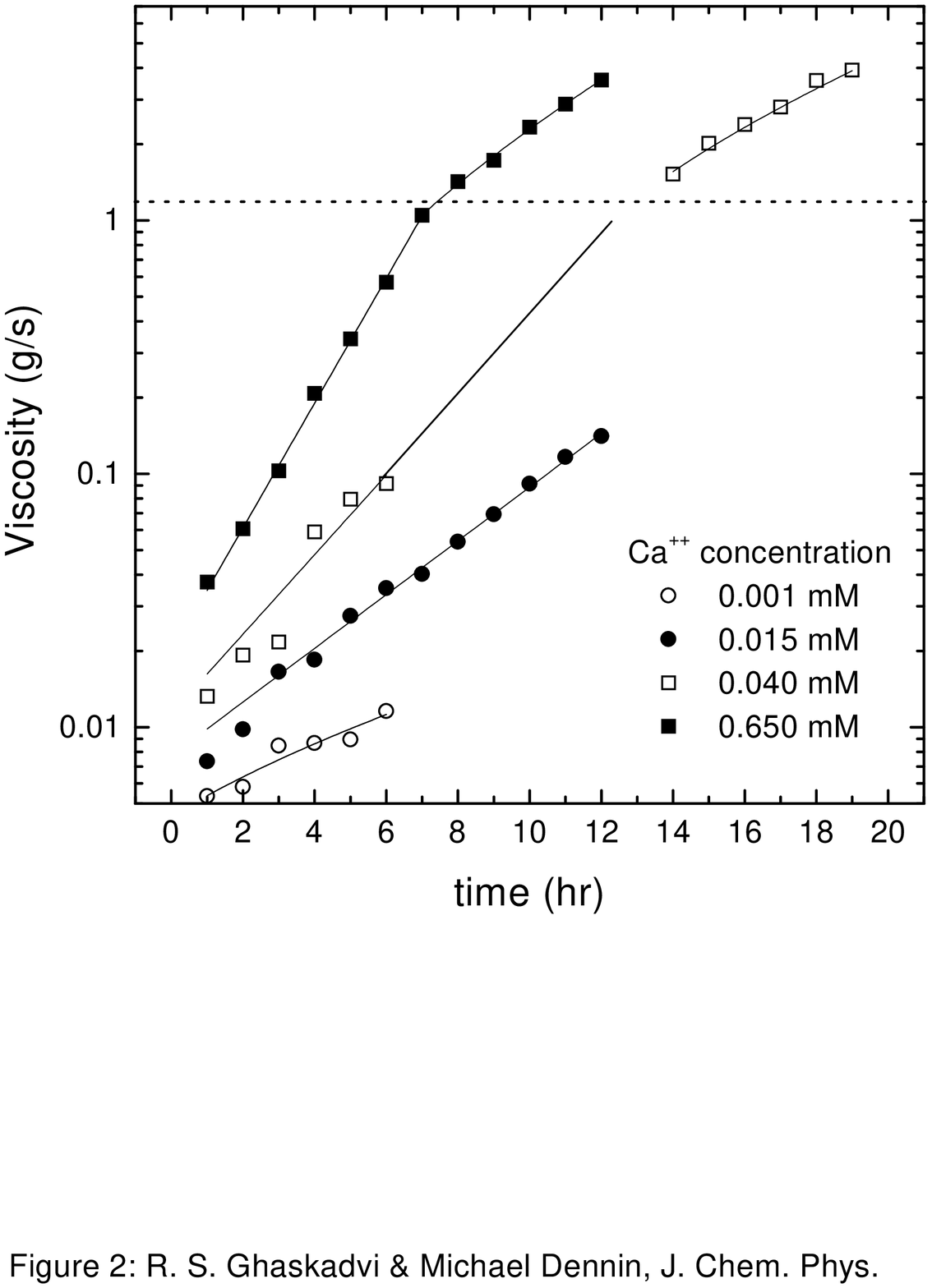}}
\caption{\label{viscotime} The viscosity of the arachidic acid monolayer as
a function of time at $22 {^\circ}{\rm C}$. The different curves correspond
to different concentrations of Ca$^{++}$ ions at pH 5.5. 
The solid lines corresponds to the least square fits to the equation 
$y = Ae{^{x/{\tau}}}$. Fit values
0.001 mM : A=0.0043 g/s, $\tau_2$=12.77 hr; 0.015 mM : A=0.0077 g/s, $\tau_2$=4.10 hr;
0.04 mM : A=0.0162 g/s, $\tau_2$=2.76 hr, $\tau_3$=8.36 hr;
0.65 mM : A=0.0197 g/s, $\tau_2$=1.76 hr, $\tau_3$=5.43 hr.}
\label{fig:viscotime}
\end{figure}

Figure~\ref{viscotime} shows the time evolution of viscosity 
(measured by the Couette flow method and henceforth referred to as $\eta$) for 
different 
concentrations of Ca$^{++}$ in the subphase. All of the measurements
were done at pH 5.5. There are two points of note. One, 
the higher the concentration, the higher the rate of viscosity rise. Secondly,
the rise in viscosity is in three parts. There is an initial jump of about
one order of magnitude within the first hour. The next two periods
are separated by $\eta$ = 1 g/s, below and above which the viscosity clearly 
rises with different slopes on the semilog plot. This indicates there are
three time constants associated with the increase. As the first data point is
taken after one hour of making the film, we cannot comment about the time
constant for the viscosity rise in the first stage, except that the upper
limit for $\tau_1$ is about 0.5 hour. It should be noted that both the 
time constants decrease with increasing concentration. 
For the film with 0.65 mM concentration of Ca$^{++}$,  $\tau_2$ = 1.76 hour and
$\tau_3$ = 5.43 hr, where $\tau_2$ and $\tau_3$ are the time constants  
for the second and the third stage respectively. The time constants for the rest
of the data are given in the figure caption.

\begin{figure}
\epsfxsize = 3.5in
\centerline{\epsffile{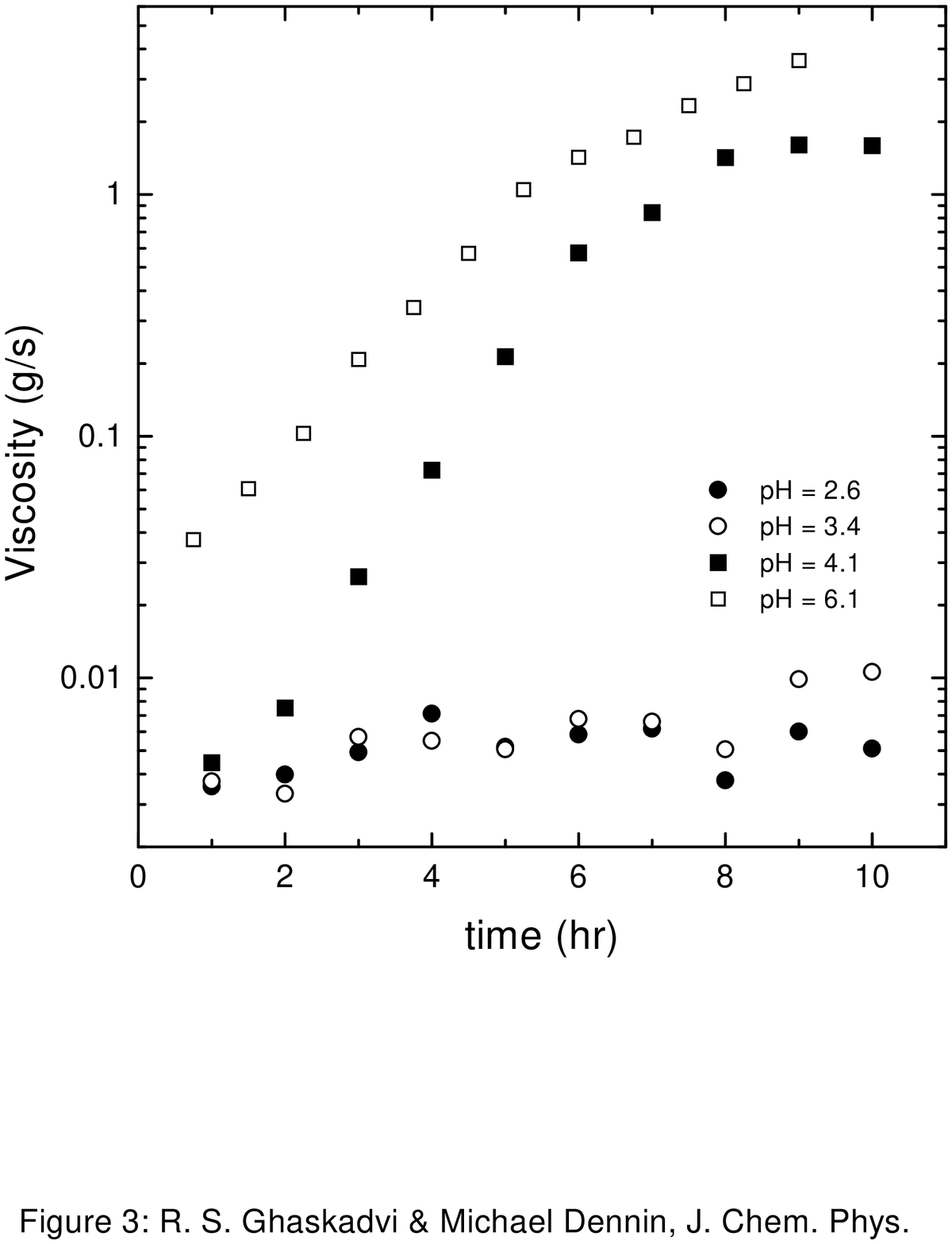}}
\caption{\label{viscoph} The viscosity of the arachidic acid monolayer as
a function of time at $22 {^\circ}{\rm C}$, $\Pi$ = 9 dyne/cm. 
The different curves correspond to different pH values of the subphase. 
The concentration of the Ca$^{++}$ ions is fixed (0.65 mM).}
\label{fig:viscoph}
\end{figure}

The three different stages of viscosity rise are also obvious in Fig.~\ref{viscoph} 
which depicts the dependence of this rise on the subphase pH. Note that below
pH = 4, the viscosity is small and almost constant. This is consistent with other 
studies where the 
isotherms were seen to remain unchanged for about the same pH. 
As the pH values are increased, both
the initial jump in viscosity and the later rates of rise increase. 

The results of the oscillatory experiment are plotted in the
Fig.~\ref{elastic}. The complex shear modulus G
(=G$^{\prime}$ + iG$^{\prime\prime}$) is known to depend 
on the strain amplitude for some Langmuir monolayers \cite{GDK92}. 
Here G$^{\prime}$ is the elastic component of the shear modulus and
G$^{\prime\prime}$ is the viscous component. For a linear
viscoelastic fluid, the relation between
G$^{\prime\prime}$ and the viscosity, $\eta$, is 
given by G$^{\prime\prime}$ = $\omega \eta$,
where $\omega$ is the oscillation frequency. In this
case we found G$^{\prime\prime}$ to be weakly and G$^{\prime}$ to be
strongly dependent on the strain amplitude. The dependence was qualitatively
the same as in Ref.~\cite{GDK92} i.e., G was constant at small amplitudes and
decreased for higher amplitudes. To ensure linear response, we measured
G at a small constant strain amplitude of about 10$^{-3}$. As with $\eta$,
G$^{\prime\prime}$ displays two distinct periods of increase after 
the first hour. However, G$^{\prime}$ rises monotonically with time. 

\begin{figure}
\epsfxsize = 3.5in
\centerline{\epsffile{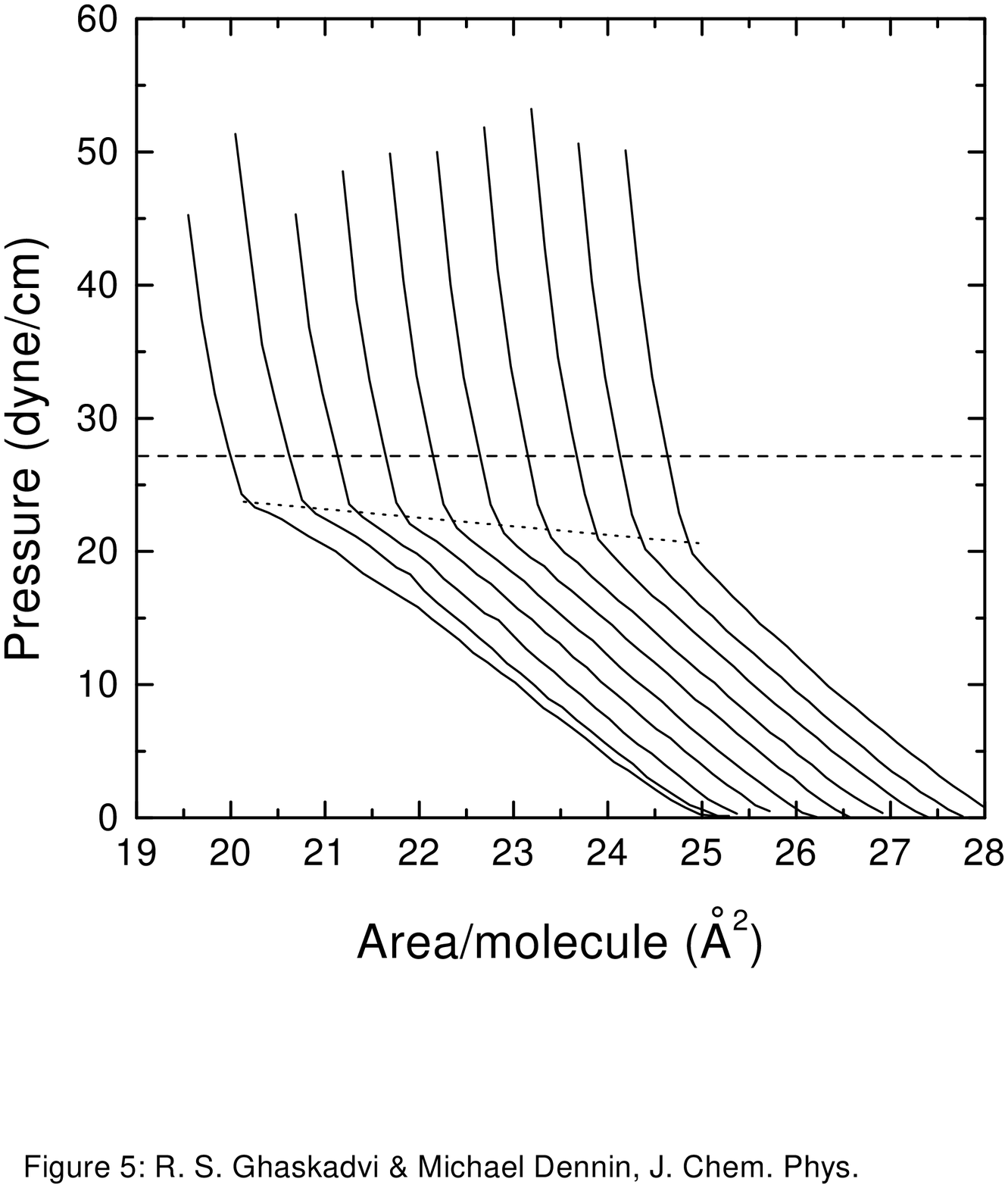}}
\caption{\label{iso} Isotherms of arachidic acid monolayer at $22 {^\circ}{\rm C}$,
subphase pH = 5.5, and subphase Ca$^{++}$ concentration of 0.65 mM.
The x-axis reading is accurate for the first isotherm only since
the rest are shifted for the sake of clarity.
The isotherms are taken one hour apart. The dotted line is drawn to guide the eye along
the kink position. The dashed line represents the pressure at which the kink occurs for
the monolayer without Ca$^{++}$ in the subphase.}
\label{fig:iso}
\end{figure}

Figure~\ref{iso} shows the variation of the arachidic acid monolayer 
isotherm as a function of time for pH = 5.5.
It must be noted that the isotherms were measured separately from
the viscosity. There might be some differences in 
the rate of Calcium attachment arising from the fact that there was no
rotation of the monolayer or generation of circular flow in the subphase.
But, the effect is bound to be minimal for two reasons. First, the flow
during viscosity measurements only occurred for roughly 10\% of the data run.
Second, there was no turbulence during the flow, so the rate of
mixing in the subphase
would not be substantially modified. Furthermore, the qualitative behavior
of G$^{\prime\prime}$ was found to be the same when measured with the rotation
as it was when measured without rotation. This confirmed that the rotation
had minimal effect on the Ca$^{++}$ binding rate.
Figure~\ref{iso} also shows the position of the kink in the isotherm that
corresponds to the 2$^{nd}$ order phase transition for
arachidic acid monolayer without Ca$^{++}$ in the subphase (horizontal dashed
line). The presence of Ca$^{++}$  does alter the isotherm
in the first hour. There is a lowering of the pressure
at which the 2$^{nd}$ order transition occurs by 
about 4 dyne/cm. This is consistent with the isotherms published in the
literature \cite{SBMZD92}. 
From the X-ray data it is known that 
this change is due to the bound calcium changing the head group interactions
so that the molecules come closer together. However, after this initial drop,
there is a slow change in the isotherm. This change corresponds to
a decrease in the transition pressure
by about 0.3 dyne/cm/hour. The kink also appears to become more rounded with time.
However, we believe that the apparent rounding is due to the high viscosity of the
film and is not a real effect.

\section{Discussion}
\label{concl}

In summary, we find that there are many effects of Ca$^{++}$ ions on the 
arachidic acid monolayer.
In the first hour, the isotherm shifts downwards in pressure by about 4 dyne/cm. Over
the next ten hours, it changes by about 3 dyne/cm. These drops are accompanied by
changes in viscosities, measured by either the rotating barrier method or the
oscillating rotor method. One can interpret the
viscosity rising during the first hour as a direct result of the change in the head group
interactions. This is consistent with the pH data. It is known that
sufficiently low pH suppresses the binding of divalent ions to the
monolayer \cite{GH94,SBMZD92,BY90,AF91},
and we observe no viscosity increase at pH 3.4 and below.
 
\begin{figure}
\epsfxsize = 3.5in
\centerline{\epsffile{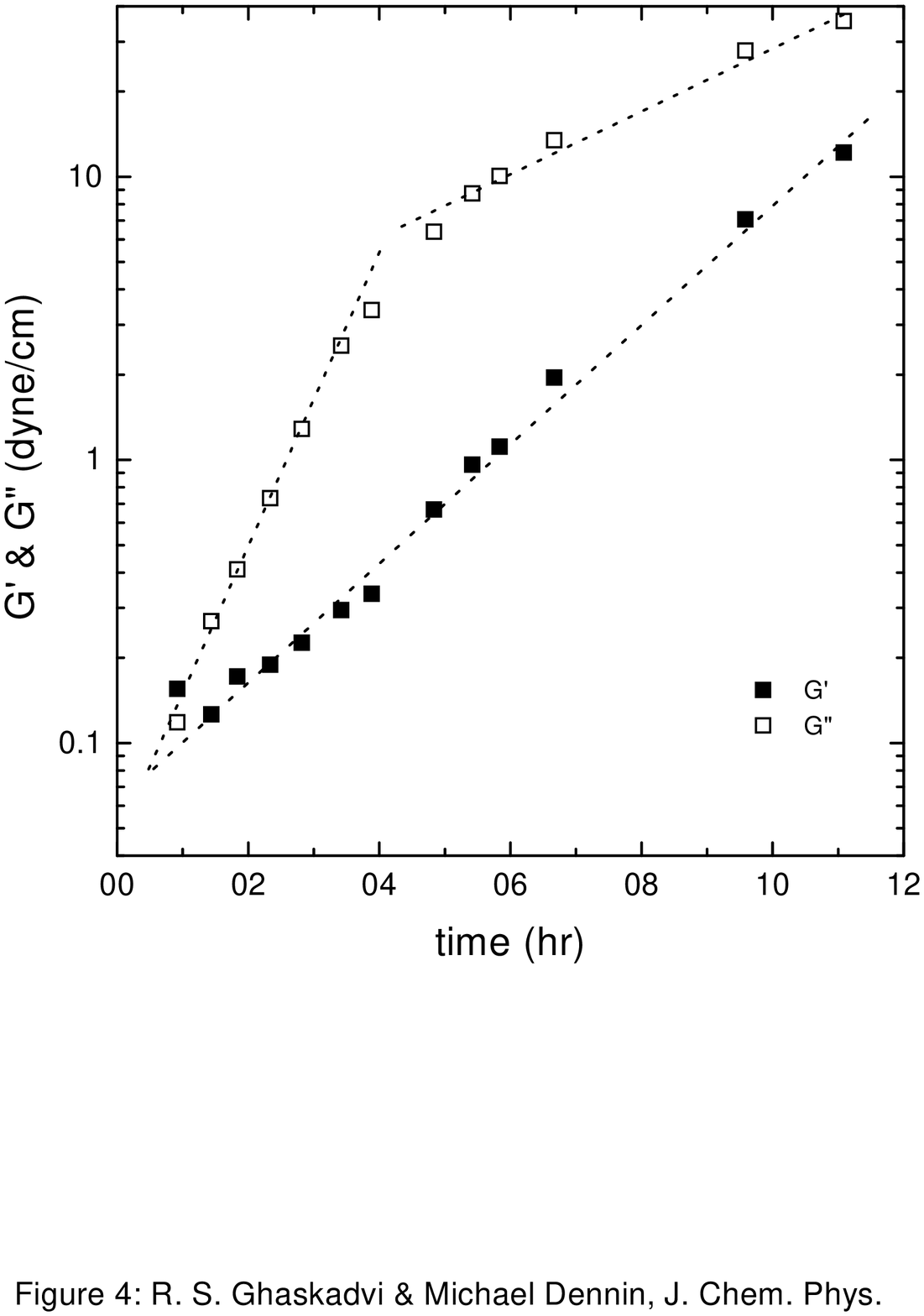}}
\caption{\label{elastic} G$^{\prime}$ and G$^{\prime \prime}$ of the arachidic acid 
monolayer as
a function of time at $22 {^\circ}{\rm C}$, $\Pi$ = 9 dyne/cm. The dotted lines are guides
to the eye.}
\label{fig:elastic}
\end{figure}

The slow rise associated with the late time evolution of the viscosity is surprising.
The 3 dyne/cm drop in the transition pressure in this period suggests a very
slow rate of Calcium ionically binding to the carboxylate. For octadecanoic acid
monolayers, IR reflection-absorption studies \cite{GH94}
have shown that near pH = 6, the Ca$^{++}$ does not bind to all the molecules but that some 
undissociated acid molecules remain in the film. 
The increase in the viscosity, taken together
with the slow change in the isotherm suggest that the same is true for the arachidic acid
and that these remaining acid molecules slowly bind with the Ca$^{++}$ ions with 
a time constant of a few hours. This is supported by the fact that the time constants 
$\tau_2$ and $\tau_3$ both decrease with increasing subphase Ca$^{++}$ concentration. 
The steady rise of  G$^{\prime}$ seen in Fig.~\ref{elastic} is consistent with this
picture; however, the existence of a single time constant needs to be explained.

The presence of two different time constants, namely $\tau_2$ and $\tau_3$, is also puzzling. 
If we accept that Ca$^{++}$ continues binding to the monolayer, then two broad possibilities
emerge: 

a. that the rate of the binding changes abruptly and this change is reflected in the viscous 
response or 

b. that the rate of binding does not change but the rate of viscosity rise with respect to 
bound site concentration varies after reaching a critical value.

At this point, it is difficult to say which
of these pictures is more accurate, but both are interesting. If the first case
is correct, it suggests interesting long-term kinetics associated with
the chemical reaction mechanism that undergo abrupt changes. If the latter reason is
correct, it suggests an interesting interplay between the microscopic structure of
the monolayer and the macroscopic viscosity.

One possible mechanism for the abrupt change in the evolution of the viscosity is
the contribution of the line tension between domains in the monolayer to the
viscosity. It is known
from foams and other complex fluids that line tension (or surface tension in
three dimensions) can substantially alter the macroscopic viscosity of a fluid.
The L$_2$ phase of arachidic acid consists of a random domain structure. Friedenberg,
et al.~\cite{FFFR96} report that for docosanoic acid monolayers in the L$_2$ phase,
domains stretched by an extensional flow do not relax back to their original shape.
This indicates that the line tension in the absence of Ca$^{++}$ is nearly zero.
Similar behavior is observed for our samples of arachidic acid. However,
with Ca$^{++}$ ions in the subphase, our BAM images show
evidence of domain relaxation. Presumably, line tension between domains
will be dominated by Ca$^{++}$ absorption at the domain boundaries. If this saturates,
the rate of change of the viscosity would be altered.
We are currently undertaking detailed studies of this behavior to probe the
impact of the line tension to the overall viscoelastic response of the monolayer and
the effect of Ca$^{++}$ ions on the line tension.

\acknowledgments

Acknowledgment is made to the donors of The Petroleum Research Fund,
administered by the ACS, for partial support of this research. Also, we would
like to thank Doug Tobias and Charles Knobler for helpful conversations.

\vskip 2.0in

\end{document}